\documentclass{sig-alternate-hotpets}

\usepackage{graphicx}
\DeclareGraphicsExtensions{.png}
\begin{document}
\title{Subtle Censorship via Adversarial Fakeness in Kyrgyzstan}

\numberofauthors{2} 
\author{
\alignauthor
Christopher Schwartz\\
       \affaddr{KU Leuven\\Centre for Ethics, Social and Political Philosophy}\\
       \email{christopher.schwartz@kuleuven.be}
\alignauthor
Rebekah Overdorf\\
       \affaddr{\'Ecole Polytechnique F\'ed\'erale de Lausanne\\Distributed Information Systems Laboratory}\\
       \email{rebekah.overdorf@epfl.ch}
}
\maketitle

\begin{abstract}
With the shift of public discourse to social media, we see simultaneously an expansion of civic engagement as the bar to enter the conversation is lowered, and the reaction by both state and non-state adversaries of free speech to silence these voices. Traditional forms of censorship struggle in this new situation to enforce the preferred narrative of those in power. Consequently, they have developed new methods for controlling the conversation that use the social media platform itself. 

Using the Central Asian republic of Kyrgyzstan as a main case study, this talk explores how this new form of ``subtle'' censorship relies on pretence and imitation, and why interdisciplinary methods of research are needed to grapple with it. We examine how ``fakeness'' in the form of fake news and profiles is used as methods of subtle censorship. 

\end{abstract}
\section{Adversarial Fakeness}

We examine fakeness at two levels: 

\begin{enumerate}
\item \emph{Fake news}, which we define as a deliberate presentation of either misleading or untrue claims as journalistic reporting.
    
\item \emph{Fake profiles}, which we define as digital entities in which either i) the persons with the alluded identities did not really create and/or control the profile, or ii) the identities are fabricated and hence cannot be attached to any real person.
\end{enumerate}

\noindent In both cases, what characterizes ``fake'' is \emph{adversarialness}: the active intention to falsify or mislead for purposes beneficial to theagent and harmful to the target. 

In concert with these cases, our preliminary results suggest the possible existence of \emph{semi-fake} profiles, in which a real user, perhaps for financial or ideological reasons, ``leases'' their identity to an adversary.

Recently, fakeness has been associated with ``hybrid warfare''~\cite{hybrid_war_example}, but this framing is too military-oriented to encompass the phenomenon. Civilian activities such as election and marketing campaigns are increasingly deploying fakeness in order to sway public opinion and consumer interest ~\emph{against} rival politicians and brands. Hence, fake news and profiles would be better conceptualized in broader terms of ``adversarial fakeness'', i.e., when malicious deception occurs as a direct action by an agent seeking to harm a target. 

Our framework enables a fuller vision of the threat the world is facing. As a form of \emph{hybrid warfare}, adversarial fakeness inflames inter-communal grievances and mobilizes populations against each other. As a form of \emph{disinformation}, it undermines the integrity of journalism, elections, and markets. As a form of \emph{censorship}, it drowns out moderate and alternative voices, and targets journalists, civil society organizations, opposition politicians and human rights activists.

\section{Subtle Censorship}

Adversarial fakeness itself can be conceptualized as ``subtle censorship.'' Insofar that censorship can be considered at root as any attempt to silence information, the phenomenon can be broadly taxonomized in a) strategic terms of withholding information, destroying information, altering or using selective information, and self-censorship, and in b) tactical terms of explicit or implicit, direct or indirect. Subtle censorship is different in that it relies primarily upon \emph{pretense} and \emph{imitation} to silence information. As a result, it obfuscates whether censorship is even happening. 

Traditional forms of censorship are attractive for scholars and activists because they imply their own solutions. In philosophical terms: \emph{if} P) there is a censor censoring, \emph{then} Q) counter-censor by circumventing (re. withholding information), archiving (re. destroying or manipulating information) or whistle-blowing or reforming (re. self-censorship). In contrast, subtle censorship eludes measurement, does not imply its own solution, and has three unusual forms:

\paragraph{Silencing through attacks}
Fake accounts are actively used to directly attack journalists and activists through social media posts and comments, the primary purpose of which is to intimidate individuals or groups into silence and submission. The example of the reporter Nedim Turfent is highly illustrative. 

In 2016, while on assignment in Turkey's Kurdish region for the pro-Kurdish Dicle News Agency, Turfent published video of soldiers threateningly standing over villagers, who were face down with their hands bound. Soon afterward, Twitter accounts that were later discovered to be linked to Turkish counter-terrorism units began taunting locals with the question, ``Where is Nedim Turfent?'' In the words of Bloomberg magazine, "The threat was clear: Give him up, or you're the next target"~\cite{turfent}. Within days, Turfent was in military custody and charged with membership in a terrorist organization. The Twitter-based manhunt abruptly ceased when one of the accounts tweeted a photograph of Turfent in custody. He remains in jail. 

Independent from the political views of Turfent, this incident illustrates the growing problem of silencing through attacks, especially with respect to journalists.

\paragraph{Silencing through noise} 
An effective method of censoring opposing viewpoints is to simply drown them out~\cite{wu2018first}. For example, automated and semi-automated fake accounts are frequently deployed to flood social networks with fake news and inflate content views, to the point that opposing perspectives appear to be in the minority or disappear down the newsfeed. As there is often an asymmetric distribution of resources favoring those already in power, this type of censorship can be quite effective and difficult to counter. So-called ``troll armies'' are the example \emph{par excellence} of this form of censorship. However, more sophisticated tactics than just sheer volume exist. For example, ``hashtag poisoning'' attacks, in which bots or fake accounts tweet a trending opposition hashtag repeatedly in order to trick Twitter into considering the hastag as spam, have been used in Mexico and elsewhere to effectively censor anti-government discussions~\cite{mexico}.

\paragraph{Silencing through misinformation}

Similar to silencing through noise, fake accounts are also used for making audiences un-receptive toward real news altogether. For example, Kyrgyzstan was destabilized in June 2010 when inter-ethnic clashes between Kyrgyz and Uzbeks were sparked by fake news about Uzbek gangs sexually assaulting Kyrgyz women and Uzbek separatists attempting to join neighboring Uzbekistan. In the years since, fake news has claimed Uzbeks are undermining the ``Kyrgyz character" of the state and are spreading radical Islam in the country. Fake news also promotes the idea that Uzbeks and the international community are engaged in an ``information war'' against the Kyrgyz nation~\cite{Meerim}. This engenders a defensive mindset throughout Kyrgyz audiences, rendering it impossible to discuss alternative views about the 2010 events.

\section{Background and Context}
Our research focuses on Kyrgyzstan, which is on the front-line of adversarial fakeness and subtle censorship. In a region notorious for authoritarianism, Kyrgyzstan boasts a generally free market and parliamentary and presidential elections held according to a schedule and rules, not the whims of ruling elites as happens among its neighbors. The quality of elections is marred by widespread vote-buying and tampering by the ruling party through state mechanisms, but the 2017 presidential election was considered genuinely unpredictable by international observers~\cite{OSCE}. In 2018, the Economist Intelligence Unit assessed Kyrgyzstan as a ``hybrid regime'' on a trajectory toward becoming a ``flawed decmocracy'' (the same category as the United States)~\cite{EIU}. 

As such, Kyrgyzstan stands out among its neighbors as having a genuinely free mediascape, with very little \emph{explicit} censorship~\cite{BBC}. The capital, Bishkek, acts as a ``haven'' for journalists from across Central Asia~\cite{haven}, and the International Exchange and Review Board has assessed Kyrgyzstan's domestic mediascape as ``near sustainable'' in its 2019 Media Sustainability Index~\cite{IREX}. Yet, self-censorship is a rampant practice among local journalists~\cite{Bakhtiyar}, partially due to a history of periodic violence against the profession, and partially due to ruling elites exploiting the weak judicial system to target the press with inflated slander lawsuits.

Throughout Central Asia, state control of ISPs is the norm, whether by overt monopolization or covert means. Kyrgyzstan not only lacks such control, but it also has ``no functional legal and technical provisions exist for shared use of existing infrastructure by ISPs, forcing them to build redundant and expensive infrastructure''~\cite{ISPs}. Although the state-owned KyrgyzTelecom remains the largest ISP within the Kyrgyz market, this situation nonetheless poses quite a hurdle to potential state surveillance and censorship.

With respect to adversarial fakeness, geostrategic issues crucial to Kyrgyzstan's sovereignty are the target of campaigns involving fake news and profiles. In 2015, fake news about an imminent nuclear war between the United States and Iran, as well as about American support for ``gay propaganda'' and Uzbek separatism, rallied public support for the pro-Russian Kyrgyz government to shut down an American military base and cancel an important 1991 bilateral treaty. Since 2018, fake news about Chinese ``imperialism'' has fuelled violent anti-Chinese demonstrations throughout the country, risking another crucial relationship. Demonstrators have called for bans on Chinese citizens from doing business in Kyrgyzstan and marrying Kyrgyz women. 
\section{New Censorship, New Methods}

As a problem to be combated, social media platforms are actively seeking solutions to adversarial fakeness and subtle censorship. For example, Facebook has publicized the fake accounts they have shut down, a majority of which are shut down at the time of the account creation~\cite{fb_tries}. The effectiveness of their methods, however, can only really be evaluated internally and these platforms lack incentives to enforce content moderation relating to fake news.

Although proposed solutions rely on machine learning methods, there are two problematic trendlines: 

\begin{enumerate}
\item \emph{No uniting framework:} Researchers focus on detecting fake news or fake profiles as interconnected yet still distinct phenomena, i.e., using one to detect the other, but not really conceptualizing them as two sides of the same coin (adversarial fakeness), and as occurring within a larger practice (i.e., subtle censorship). \item \emph{No human focus:} Related to the lack of a uniting framework, researchers also tend to focus on the fake news propagation role of fake profiles, rather than on the \emph{re-propagation} role of audiences. 
\end{enumerate}

Our project in Kyrgyzstan broadens this research by combining machine learning, philosophy, and journalism. With our partners the Civic Initiative for Internet Policy, a Kyrgyz civil society organization, we have assembled a team of journalists and researchers in Bishkek who are assisting us in the following.  
\begin{enumerate}
\item Establishing ground truth by manually identifying a) fake accounts that seed fake news, and b) human beings that serve as re-propagators. 
\item After crawlers derive social graphs on both populations, deriving feature sets, then training classifiers upon these features. 
\item Validating results journalistically by a) utilizing context knowledge to critically analyze the fake news samples and b) endeavoring to confirm claims made by various sources regarding the origins of the examined fake profiles.
\end{enumerate}

\section*{Acknowledgements} This project is funded by the Open Technology Fund's Information Controls Fellowship. 


\bibliographystyle{abbrv}
\bibliography{citations}  

\balancecolumns
\end{document}